\documentclass[aps,prl,twocolumn,superscriptaddress,groupedaddress]{revtex4}

\usepackage{graphicx}
\usepackage{color}
\usepackage{dcolumn}  
\usepackage{bm}       
\usepackage{amssymb}
\usepackage{amsmath}
\usepackage{lipsum}
\usepackage{enumitem} 

\usepackage[normalem]{ulem}

\newcommand{\Pmn}{P_m(n)}
\newcommand{\Pmz}{P_m(z)}
\newcommand{\mn}{\langle m(n) \rangle}
\newcommand{\mz}{\langle m(z) \rangle}

\newcommand{\mmz}{\langle m^2(z) \rangle}
\newcommand{\mkn}{\langle m^k(n) \rangle}
\newcommand{\mkz}{\langle m^k(z) \rangle}
\newcommand{\kappakn}{\kappa_k(n)}

\newcommand{\leadingM}{\langle m^k(n) \rangle_a}
\newcommand{\leadingK}{\kappa_k^{(a)}(n)}
\newcommand{\subleadingM}{\langle m^k(n) \rangle_s}
\newcommand{\subleadingK}{ \kappa_k^{(s)}(n) }

\begin{document}


\title{Zero-Crossing Statistics for Non-Markovian Time Series}
\author{Markus~Nyberg}
\affiliation{Integrated Science Lab, Department of Physics, Ume\r{a} University, SE-901 87 Ume\r{a}, Sweden}
\author{Ludvig~Lizana} \affiliation{Integrated Science Lab, Department of Physics, Ume\r{a} University, SE-901 87 Ume\r{a}, Sweden}
\author{Tobias~Ambj\"{o}rnsson} \email{tobias.ambjornsson@thep.lu.se}  \affiliation{Department of Astronomy and Theoretical Physics, Lund University, S\"{o}lvegatan 14A, SE-223 62 Lund, Sweden}

\date{\today}


\begin{abstract}

In applications spaning from image analysis and speech recognition, to energy
dissipation in turbulence  and time-to failure of fatigued materials,
researchers and engineers want to calculate how often a stochastic observable
crosses a specific level, such as zero. At first glance this problem looks
simple, but  it is in fact theoretically very challenging. And therefore, few
exact results exist. One exception is the celebrated Rice formula that gives
the mean number of zero-crossings in a fixed time interval of a zero-mean
Gaussian stationary processes. In this study we use the so-called Independent
Interval Approximation to go beyond Rice's result and derive analytic
expressions for all higher-order zero-crossing cumulants and moments. Our results agrees well with simulations for the non-Markovian autoregressive model.
\end{abstract}

\maketitle

\textbf{\emph{Introduction}.} 
Imagine a  stationary time series that repeatedly crosses through its mean, say zero. Because it is stationary, the crossings occur with a constant rate $r$, which means that the ensemble-averaged number of zero-crossings $\mn$ during a fixed time interval $n$ is
\begin{equation} \label{eq:rice0}
\langle m(n) \rangle=nr .
\end{equation}
Although a simple formula, it is a challenge to derive a closed-form
expression for $r$ in terms of the statistical properties of the time series. One exception is Gaussian stationary time series $X(0),\ldots,X(n)$,  where in 1945 Stephen Rice found that $r = (1/\pi)\times\arccos[\langle X(k)X(k+1) \rangle/\langle X(0)^2\rangle]$
\cite{rice1944mathematical}. 

To honour his work, Eq. \eqref{eq:rice0} is referred to as  the Rice formula, and has over the last seven decades been applied to a  variety of scientific and engineering problems 
For example to calculate energy dissipation in turbulent flows \cite{poggi2010evaluation},  time-to failure from material fatigue in  structures exposed to random loads such as wind \cite{mrvsnik2013frequency}, and to estimate occurrences of high ocean waves \cite{rychlik2000some}. However, apart from Rice's formula  and a few special cases \cite{majumdar2002statistics,ehrhardt2004persistence}, there are no general analytical results for the zero-crossing moments and cumulants.  This limits the  formula's predictive power because we cannot reliably calculate error estimates and confidence intervals. Here, we narrow this knowledge gap.

To estimate the zero-crossing cumulants, let us briefly assume that the zero-crossing process is described by flipping a skewed coin; the probability of "heads" is $r$, and "tail" is $1-r$. If we get heads, we cross zero between times $n'-1$ and $n'$,  and if we get tail we do not cross. If the coin flips are independent, the distribution of $m$ zero-crossings during  $n$ time steps  is given by the binomial distribution
 $\Pmn ={n \choose m} r^m(1-r)^{m-n}$.
And just as Rice formula, this process gives $\langle m(n) \rangle = nr$. Moreover,  any
cumulant $\kappakn$  is linear in $n$
\begin{equation} \label{eq:kappaAndM}
\kappakn = K_k n, 
\end{equation}
where the first three pre-factors are  $K_1=r$, $K_2 =r(1-r)$ and $K_3=r(1-r)(1-2r)$. 

However, the zero-crossing statistics for a general stationary process is not that of a coin-flip   process -- the probability that  $X(n')$ crosses zero between times $n'-1$ and $n'$ typically depends on the trajectory $X(n'-1),X(n'-2),\ldots$ that leads up to $n'$. How does
Eq. \eqref{eq:kappaAndM} change under these conditions? Using a generalisation of the so-called Independent
Interval Approximation (IIA) \cite{nyberg2016simple}, we find that the scaling with $n$ holds for  a  general Gaussian stationary time series. We also derive expressions for the modified pre-factors $K_k$ in terms of the process' autocorrelation function, and  a recursion formula for the sub-linear contributions to $\kappakn$.

\textbf{\emph{Zero--crossing probability density}.}
To quantify the zero-crossing statistics of a time series $X(0),\ldots X(n)$ within IIA, 
we will need the probability  $\omega(n|x_0)$ that $X(n)<0$ given that $x_0=X(0)>0$. If  $X(n)$ is  Gaussian with mean $\mu(n)$ and variance $\sigma^2(n)$, we have $\omega(n|x_0)={\rm erfc}(\mu(n)/\sqrt{2\sigma^2(n)}/2)$. In terms of the covariance $\gamma(n)=\langle x(n+k)x(k) \rangle$ and the normalised autocorrelation function $A(n) = \gamma(n)/\gamma(0)$, we may write $\mu(n)=X(0)A(n)$  and $\sigma^2(n)=\gamma(0)\left(1-A(n)^2\right)$ \cite{brockwell2006introduction}.
Averaging $\omega(n|x_0)$ over  the equilibrium density $\varrho(x_0)$,
$\omega(n) = \int_{-\infty}^\infty \omega(n|x_0>0) \varrho(x_0) dx_0$, leads to (see Appendix)
\begin{equation} \label{eq:omegan}
\omega(n) =  \frac 1 2  + \frac{\arcsin A(n)}{\pi}.
\end{equation}
This is the probability that $X(n)$ is below zero given the thermalised initial position $x_0>0$. Our expressions for the cumulants $\kappakn$ and
moments $\mkn $ $(k=1,2,\ldots$) will be expressed in terms of $\omega(n)$.

\begin{figure}[] 
	\centering
	\includegraphics[width=0.9\columnwidth]{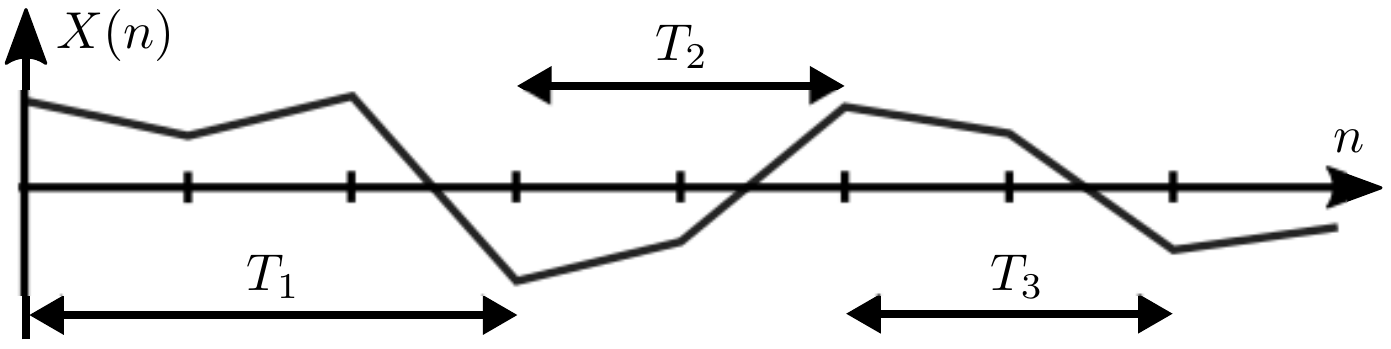}
	\caption{Stochastic time series $X(n)$ of a continuous value $x(n)$ as a function of the discrete time $n$. Time intervals $T_1,T_2,\ldots$ denote times spent above and below the origin.  \label{fig:x(n)}}
\end{figure}

Now we derive an expression for the zero-crossing probability $\Pmn$ in terms of $\omega(n)$ using the IIA framework.  To that end, consider  $X(n)$  illustrated in Fig.~\ref{fig:x(n)}. As time progresses, $X(n)$ repeatedly goes through zero (its mean), and we may divide the total observation time into intervals  where $X(n)>0$ ($T_1,T_3,\ldots$)  and $X(n)<0$ ($T_2,T_4,\ldots$). Now, the IIA's main assumption is that the length of these intervals are independent. The probability density function  of the first time interval (the first-passage density) $T_1$, is denoted by $\rho(T_1)$. Subsequent interval lengths  ($T_2,T_3,T_4,\ldots$) are random numbers, $n$, described by the first-return density $\psi(n)$. 

Using the IIA, we can express $\Pmn$ in terms of $\psi(n)$ and  $\rho(n)$. To see this, consider the probability to make a single zero-crossing up to $n$. This is the probability that $X$ first crossed at $T_j$ multiplied by the probability that it  did not cross between $n-T_j$ and $n$. That is $P_1(n) =  \sum_{j=1}^n \rho(T_j)(1-\sum_{j'=j}^n\psi(T_{j'}))$. Generalising this argument to $m$ crossings yields \cite{mcfadden1958axis,sire2007probability}
\begin{equation} \label{eq:pmn11}
\begin{aligned}
P_{m}(n) &= \sum_{j_1=0}^{n}\rho(T_{j_1})\sum_{j_2=j_1}^{n}\psi(T_{j_2}-T_{j_1})\sum_{j_3=j_2}^{n}\psi(T_{j_3}-T_{j_2}) \\
&\cdots \sum_{j_{m}=j_{m-1}}^{n}\psi(T_{j_{m}}-T_{j_{m-1}})\left(1-\sum_{j'=j_m}^n\psi(T_{j'})\right).
\end{aligned}
\end{equation}

To derive Eq. (\ref{eq:pmn11}), we assumed that $\psi(n)$  is the first-return probability to zero, and that it is the same for all crossings. While this is true in continuous-time, it is an approximation in discrete-time because it is rare that the process ends up exactly at zero (it rather over-shoots zero, see Fig. \ref{fig:x(n)}). However,  as discussed in  \cite{nyberg2016simple}, this only weakly affect our IIA results.

To simplify Eq. \eqref{eq:pmn11}  we take the  $z$-transform  $f(z)=\sum_{n=0}^{\infty}f(n)z^{-n}$. This gives
\begin{equation} \label{eq:pmn2}
P_{m}(z) = \rho(z)\psi(z)^{m-1}\frac{z\left[ 1-\psi(z)  \right]}{z-1}.
\end{equation}
Next we eliminate  $\psi(z)$ and $\rho(z)$ to express $\Pmz$ solely in terms of $\omega(z)$. First, we demand that our equations satisfy the  Rice formula $\sum_{m=0}^\infty m \Pmn = nr$. This gives
\begin{equation}\label{eq:cond1}
 \rho(z) = r \frac{1-\psi(z)}{z-1}.
\end{equation}
Second,  we use that the probability for $X(n)$ to be below zero given that $X(0)$ started somewhere above zero, i.e. $\omega(n)$, is the same as the probability that $X(n)$ crossed zero an odd number of times (Fig. \ref{fig:x(n)}). That is  $\omega(n)=\sum_{m=1}^{\infty}P_{2m-1}(n)$.  Using Eq. \eqref{eq:pmn2}  in the sum gives
\begin{equation} \label{eq:cond2}
\omega(z) =  \frac{z\rho(z)}{z-1}\frac{1}{1+\psi(z)},
\end{equation}
and combining Eqs. \eqref{eq:cond1} and \eqref{eq:cond2} leads to
\begin{equation}\label{eq:Pmz_final}
P_m(z)= \frac{4 rz(z-1)^2 \omega(z)^2 }{r^2 z^2-(z-1)^4\omega(z) ^2} \left(\frac{2 r z}{r z+(z-1)^2\omega(z)  }-1\right)^m.
\end{equation}
This is an exact expression for the zero-crossing probability density within the IIA.

%
 
\textbf{\emph{Zero-crossing cumulants}.} 
To calculate the cumulants $\kappakn$, we first use Eq. \eqref{eq:Pmz_final} to get the moments  $\mkz = \sum_{m=0}^{\infty}m^kP_m(z) $. The first four  are
\begin{eqnarray}  \label{eq:moments}
&&\mz = \frac{rz}{(z-1)^2}, \ \ \ \mmz = \frac{2 r^2 z}{(z-1)^3q(z)},  
\nonumber\\
&&\langle m^3(z)\rangle =\frac{6r^3z}{(z-1)^4q(z)^2}-\frac{rz}{2(z-1)^2}, \\
&&\langle m^4(z)\rangle =\frac{24r^4z}{(z-1)^5q(z)^3}-\frac{4r^2z}{(z-1)^3q(z)},
\nonumber
\end{eqnarray}
where  
\begin{equation}\label{eq:q_z}
q(z) = \frac {2(z-1)\omega(z)} z.
\end{equation}
To invert these we must resolve the inverse $ \oint_C \mkz  z^{n-1} dz/(2\pi
i)$  \cite{debnath2014integral}. Here the curve $C$ is outside the unit circle because the magnitude of the largest pole is at most unity for stationary processes \cite{lindgren2012stationary}. From Eq. \eqref{eq:moments}, we see that there is a pole at $z=1$, and possibly others that satisfy $q(z)=0$. To find out, we use  Eqs. \eqref{eq:omegan} and \eqref{eq:q_z} to write $q(z) = 1-[2(z-1)/(\pi z)]  \sum_{i=0}^\infty  \arcsin [A(i)]z^{-i}$. From this we see that $z=1$ is not a root to $q(z)$ because $q(1) = 1$. This means that all zeros to $q(z)$ must be strictly inside the unit circle. This allows us to decompose $\mkn$ into a leading and a sub-leading part, $\leadingM$ and $\subleadingM$ respectively,  that dominate the dynamics for asymptotically large 
and small $n$:
\begin{eqnarray}\label{eq:momentDecomp}
\mkn &=& \leadingM + \subleadingM,
\end{eqnarray} 
This argument also holds for the cumulants $\kappakn = \leadingK + \subleadingK$. Below we calculate the leading  term from the residue at $z=1$, and derive a recursion relation for the sub-leading part.

\textbf{\emph{Asymptotic behaviours}.} 
We calculate $\leadingM$ as
\begin{equation} \label{eq:residue+asymptotic}
\leadingM ={\rm Res}_{z\to 1}\mkz z^{n-1}.
\end{equation}
Although in closed form, the formulas for $\leadingM$ are tedious (see
Supplementary material), but due to cancellations, the cumulant
expressions become much simpler. Besides the Rice formula for $\kappa_1(n)$,
we find to leading order in $n$ that
\begin{equation}\label{eq:kappa2}
\kappa_2^{(a)}(n) \simeq n r^2 \left(-2 q'(1)-1\right) 
\end{equation}
\begin{equation}\label{eq:kappa3}
\begin{aligned}
\kappa_3^{(a)}(n) \simeq n \left(r^3 \left(-3 q''(1)+12 q'(1)^2+6 q'(1)+2\right)-\frac{r}{2}\right)
\end{aligned}
\end{equation}
\begin{equation}
\begin{aligned}
&\kappa_4^{(a)}(n) \simeq  2 n r^2 \Big(2 q'(1)-r^2 (2 q^{(3)}(1)-6 q''(1) \\
&+60 q'(1)^3+30 q'(1)^2-6 q'(1) (5 q''(1)-2)+3)+1\Big)
\end{aligned}
\end{equation}
where $q^{(n)}(1)$ is the $n$:th derivative of $q(z)$ evaluated at $z=1$. Here we see that $\kappa_k(n)\simeq K_k n$ as in Eq. \eqref{eq:kappaAndM},
but with non-trivial pre-factors. 

\textbf{\emph{Sub-leading behaviours}.} 
It is not convenient to use residue analysis to calculate the sub-leading contribution because $q(z)$ is not on a suitable analytical form. We therefore  derived recursive relations for $\mkn$ (see Appendix). For $ k=2$ we have
\begin{eqnarray}\label{eq:second_moment_recur}
&&\langle m^2 (n) \rangle_s={\cal G}^{(2)}(n)-\frac{1}{r}\sum_{k=0}^{n-1}\langle m^2 (k) \rangle_s\omega(n-k+1)),\nonumber\\
&&{\cal G}^{(2)}(n) =  \frac{r}{6}n(n+1)(n+2)\nonumber\\
&&-r^2\left(n(n-1)-2 n q^{(1)}(1)+2 q^{(1)}(1)^2-q^{(2)}(1)\right)\nonumber\\
&&-r\sum_{k=0}^{n-1}\left(k(k-1)-2 k q^{(1)}(1)+2 q^{(1)}(1)^2-q^{(2)}(1)\right)\nonumber\\
&&\ \ \ \times\omega(n-k+1).
\end{eqnarray}
We give the $k=3$ - case explicitly in the Appendix. In fact, it is possible to invert all moments in Eq. \eqref{eq:moments} recursively, but this is not convenient for asymptotic analysis.


\begin{figure}[] 
 \centering
 \includegraphics[width=\columnwidth]{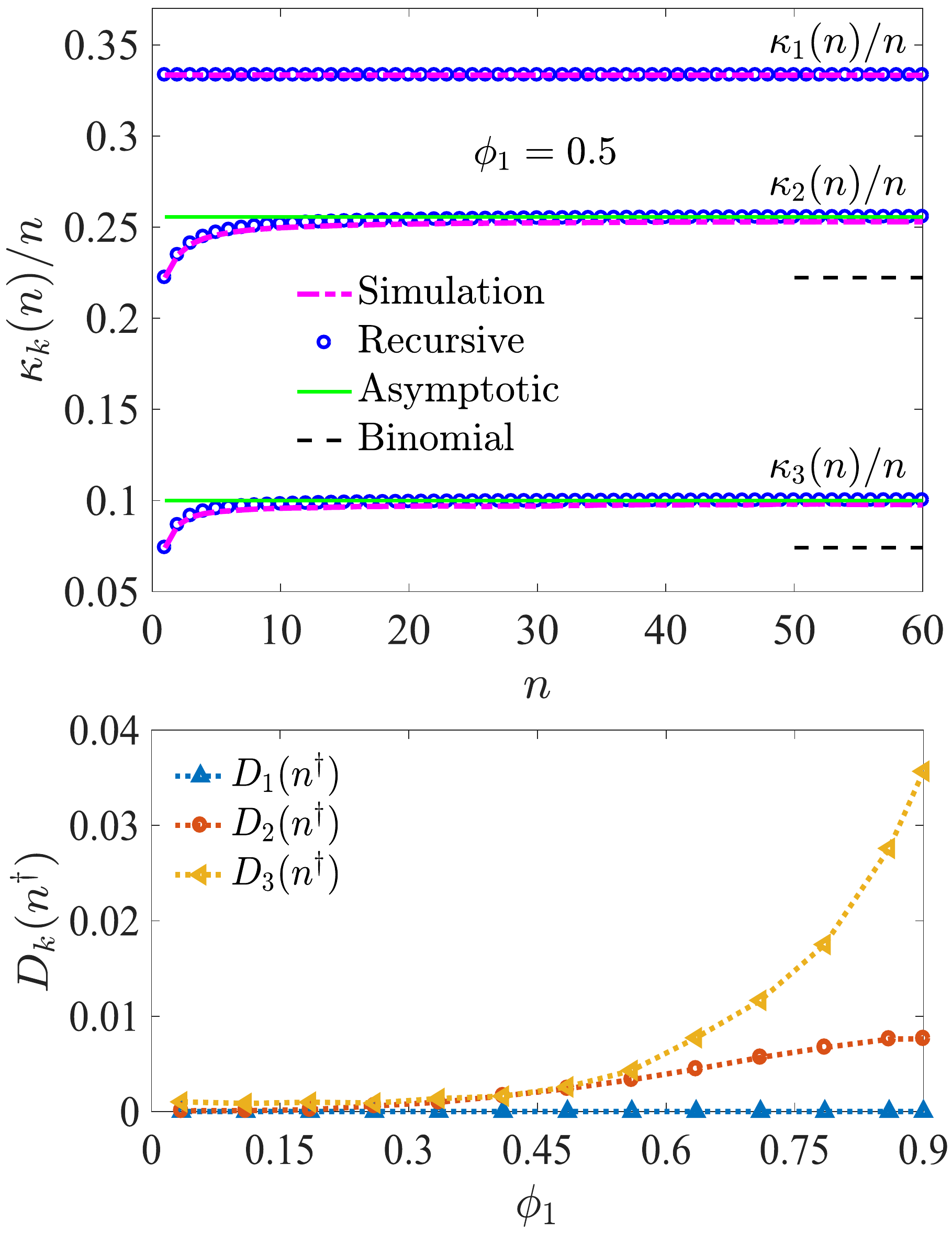}
 \caption{(top) Zero-crossing cumulants $\kappakn$ ($k=1,2,3)$ for the first-order autoregressive model $x(n) = 0.5x(n-1)+\eta(n)$. (bottom)  The cumulants' absolute difference $D_k(n^{\dagger})$ between simulations and asymptotic results at time $n^\dagger=150$. Simulations are averaged over $10^7$ realizations. \label{fig:AR1}}
\end{figure}

\textbf{\emph{Numerical results}.} 
Here we apply our expressions to the autoregressive model of order $p$
\cite{lindgren2012stationary}, denoted  as AR($p$), that often appears in time
series analysis \cite{brockwell2006introduction,lindgren2012stationary}. The
key parameter $p$ determines the time span of the process memory (for $p=1$ the model is Markovian), and the general equation of motion for AR($p$) is ($n\geq p$)
\begin{eqnarray} \label{eq:arma1}
x(n) = \phi_1x(n-1)+\ldots +\phi_p x(n-p)+\eta(n),
\end{eqnarray}
where $\phi_1\ldots \phi_p $ are constants, $\eta(n)$ is (Kronecker) delta-correlated noise,  and $\langle \eta(n_1)\eta(n_2)\rangle= \Sigma^2 \delta_{n_1,n_2}$ with amplitude $\Sigma$ (below we set $\Sigma=1$).
To quantify how well our expressions match real processes we compare them to
simulations. We use a computationally fast method from
\cite{wood1994simulation} that uses the covariance $\gamma(n)$ as input,  rather than simulating Eq. \eqref{eq:arma1} directly. In the Appendix we give $\gamma(n)$ in terms of $\phi_1\ldots \phi_p $ for the relevant processes.

{\it Example 1: AR(1)}. Also known as the discrete-time Ornstein-Uhlenbeck
process, AR(1) is the simplest autoregressive model. It has the
autocorrelation $A_1(n)=\phi_1^n$.  In Fig. \ref{fig:AR1}
(top), we show how the first three cumulants grow with $n$ when
$\phi_1=0.5$. We compare the asymptotic results
$\leadingK$, with results from the recursion relations which are exact
within IIA (see Appendix) and results based on simulated AR(1) trajectories. To better illustrate the long time behaviour, we scale $\leadingK$ by $n$. For comparison, we also included the cumulants for the Binomial distribution [Eq. \eqref{eq:kappaAndM}].

In Fig \ref{fig:AR1} we see that our expressions and simulations agree very well
when  $\phi_1=0.5$. The agreement is substantially better than  the binomial
approximation. To further quantify how much  simulations and IIA results deviate for other values of $\phi_1$, we varied $\phi_1$ and made similar plots as Fig. \ref{fig:AR1} (top). From these we calculated the the absolute error $D_k(n^\dagger)$ at some large time point $n^\dagger$, between simulated and asymptotic values, that is $D_k(n^\dagger) = |\kappa_k^{(\rm Sim.)}(n^\dagger)-\kappa_k^{(\rm a)}(n^\dagger)|/n^\dagger$. In Fig. \ref{fig:AR1} (bottom) we see that the errors are small, especially when $\phi$ goes to zero (as it should because the correlations between time points vanishes as $\phi_1\rightarrow 0$).

{\it Example 2: AR(3)}. Next we study the non-Markovian three step process $x(n) = \phi_1x(n-1)+\phi_2x(n-2)+\phi_3x(n-3)+ \eta(n)$, with autocorrelation function  
\begin{equation} \label{eq:A3}
A_3(n)=C_1\lambda_1^{n}+C_2\lambda_2^{n}+C_3\lambda_3^{n},
\end{equation}
where $C_i$ are constants that depends on $\lambda_1$, $\lambda_2$ and $\lambda_3$ which in turn depends on $\phi_1,\phi_2$ and $\phi_3$. We give these explicitly in the Appendix. 

Figure \ref{fig:AR3} is the same as Fig. \ref{fig:AR1} but for AR(3). In the top panel we put $\lambda_1=0.5$, $\lambda_2=0.2$ and $\lambda_3=0.1$. Again we see good correspondence between simulations and analytical results but the deviations grow for the higher-order cumulants. In the lower panel we varied $\lambda_1$ while fixing $\lambda_2=0.2$ and $\lambda_3=0.1$ Although larger than for AR(1), the absolute errors for AR(3) are small.

\begin{figure}[] 
 \centering
 \includegraphics[width=\columnwidth]{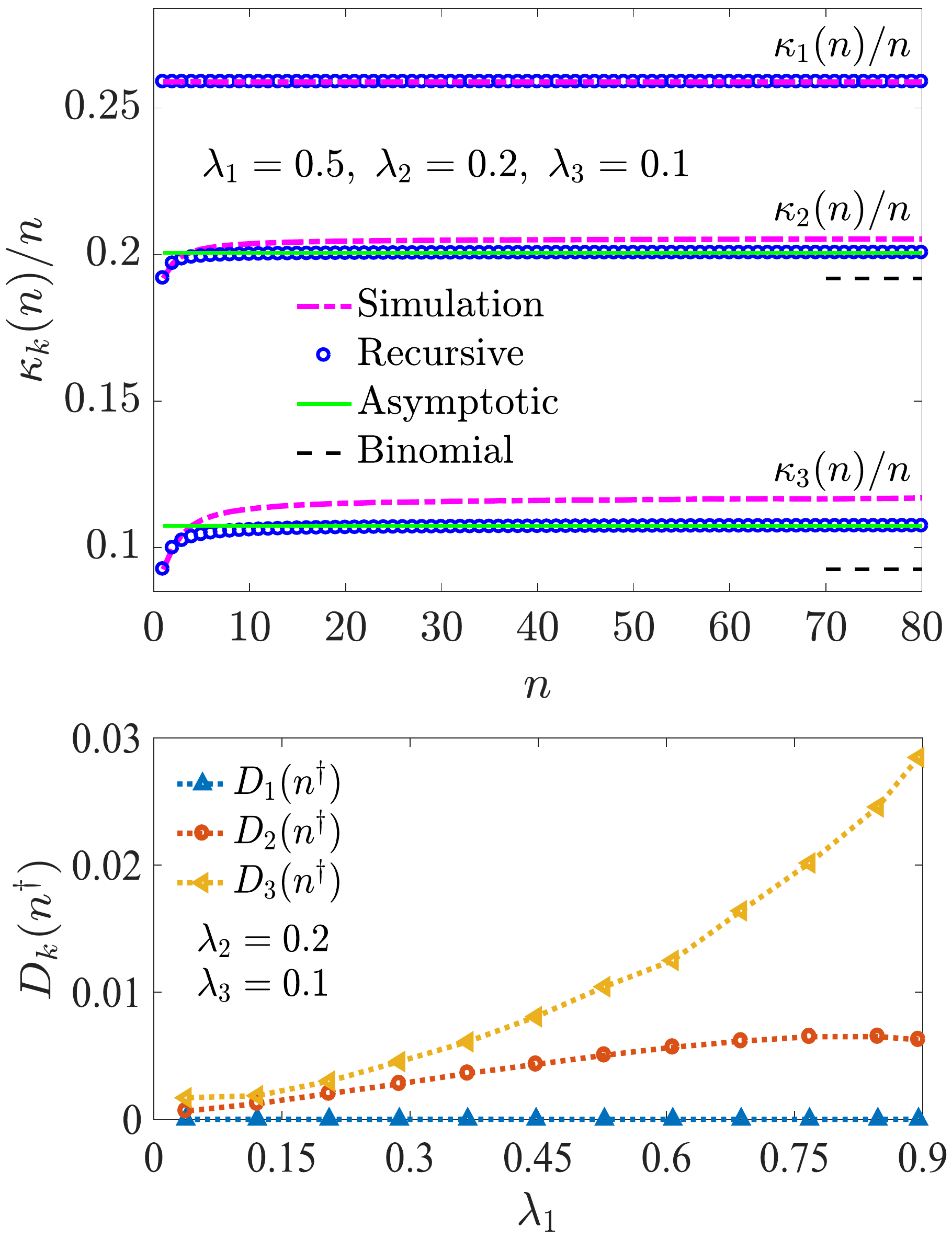}
 \caption{(top) Zero-crossing cumulants $\kappakn$ ($k=1,2,3)$ for the non-Markovian third-order autoregressive model, AR(3), with $\lambda_1=0.5$, $\lambda_2=0.2$ and $\lambda_3=0.1$. (bottom)  Absolute error $D_k(n^{\dagger}=300)$ between simulations and asymptotic results  when $\lambda_2=0.2$ and $\lambda_3=0.1$.  As $\lambda_1$ grows from $0$ to $0.9$, $\phi_1,\phi_2$ and $\phi_3$ grows linearly: $\phi_1:0.3\to 1.2$, $\phi_2:-0.29\to-0.02$, and $\phi_3:0\to 0.018$. We  averaged over $2\times10^7$ trajectories.
 \label{fig:AR3}}
\end{figure}

\begin{figure}[] 
 \centering
 \includegraphics[width=\columnwidth]{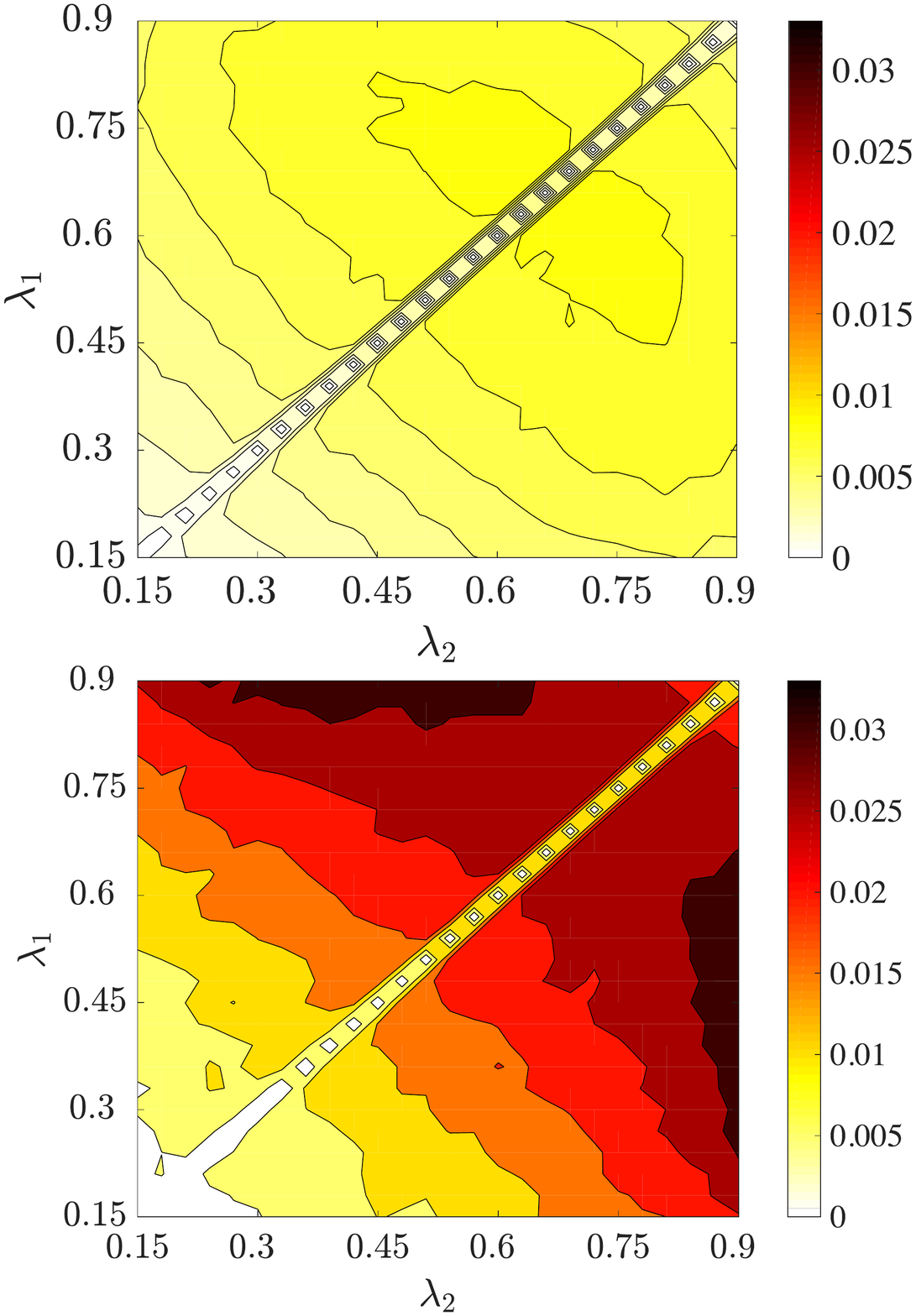}
 \caption{Absolute error between simulations and asymptotic analytical results of the cumulants' linear coefficient for the autoregressive model AR(11). Top: $D_2(n^{\dagger}=300)$. Bottom: $D_3(n^{\dagger}=300)$. Simulations are averaged over $10^7$ realizations. \label{fig:AR11}}
\end{figure}

{\it Example 3: AR(p).}
The IIA is handy to derive closed form expressions, but it is an approximation that ought to break down when correlations span several consecutive crossing intervals. We therefore ask: under what conditions can our method make accurate predictions for a general AR($p$) process? To answer this, we consider  AR(11)  that has the autocorrelation function $A_{11}(n)=C_1\lambda_1^{n}+C_2\lambda_2^{n}+\cdots+C_{11}\lambda_{11}^{n}$ (see Appendix for details). Guided by the results for AR(1) and AR(3), we anticipate that the IIA will work well for AR($p$) as long as the largest of $\lambda_1,\ldots,\lambda_p$ is much larger than the 2nd largest and so on. To test this, we set $\lambda_3=0.1$, $\lambda_k=\lambda_{k-1}/2$ for $k=4,\ldots,11$ and varied $\lambda_1$ and $\lambda_2$ from 0.15 to 0.9. In Fig. \ref{fig:AR11} we plot the absolute errors $D_2(n^{\dagger})$ and $D_3(n^{\dagger})$ as heat maps. As anticipated, we get accurate predictions when $\lambda_1\gg \lambda_2$ (and  $\lambda_1 \ll \lambda_2$ because of symmetry). In particular, we note that the errors are similar to AR(1) and AR(3) ($D_2(n^{\dagger})\lesssim0.005$ and $D_3(n^{\dagger})\lesssim0.03$) even when $\lambda_1\approx\lambda_2$. 

\textbf{\emph{Discussion}.} 
The conditions for  $\lambda_3,\ldots\lambda_{11}$ in the AR(11) example ensure that the core assumptions of the IIA are not too violated. In other words, memory effects do not extend over several zero-crossings. This means that the IIA cannot in its present form handle strongly correlated processes, such as fractional Gaussian noise where $A(n)\sim n^{-\beta}$ with $\beta\in (0,2)$. Generalising IIA to these processes is a big challenge that goes beyond the scope of this work. 

The autoregressive model AR$(p)$ that we  used as case study can be seen as a discrete-time version of $d^px(t)/dt^p=\eta(t)$ in logarithmic time \cite{bray2013persistence}. In \cite{ehrhardt2004persistence}, the authors calculated the zero-crossing distribution $P_m(n)$ for this process as a series expansion in $A(n)$  up to 10th order with help of Mathematica. As an example, they compared their results for  $x(n)=(\eta(n)+\eta(n-1))/\sqrt{2}$ with the asymptotically exact expression \cite{majumdar2002statistics} and found good correspondence. However, the method has convergence problems when the crossings are too frequent ($m/n \lesssim0.73$). In addition to $P_m(n)$, they also calculated the variance $\kappa_2(n)$ up to 14th order in a similar way. They achieve best results for the AR(1) process, but again there are  convergence problems  (when $\phi\gtrsim 0.75$). Although in principle exact, their method has a few shortcomings that we do not share in our work. And, our approach provides  closed-form, arguably simpler,  expressions.

Finally, even though we focused on time-series, where the time between
recordings is constant, all our results can be extended to continuous time
\cite{abrahams1986survey,blake1973level}. If we put $t_n=n\epsilon$, with
$\epsilon$ as the time between events, we get the continuum case when
$\epsilon\rightarrow 0$, $n\rightarrow \infty$ with $t=n\epsilon$ held
fixed. Also, in this limit we must replace the the zero-crossing rate with
$\sqrt{-A''(t_0)}/\pi$ \cite{bray2013persistence}. Overall, this means that
our expressions for $\kappakn$ are approximations to the continuum limit for
smooth processes (were also the overshooting effect disappear). However, we
point out that some results for this limit already exists. For example, the
authors of \cite{blake1973level,smith2008fluctuations} show that the variance,
$\kappa_2(n)$, of any continuous Gaussian zero-mean process can be expressed
as an integral that, however, must be evaluated numerically.

\textbf{\emph{Summary and outlook}.} 
How many times does a Gaussian stationary  process cross zero in a fixed time interval? And how much does that number fluctuate? To answer this, we used a generalisation of the Independent Interval Approximation. With this  we calculated all zero-crossing cumulants in terms of the process' autocorrelation function. As our work extends the heavily used Rice formula that only gives the average, we look forward to application of our methods and results in zero-crossing-based problems such as signal analysis \cite{shenoy2015spectral}, wireless communication \cite{iskander2002analytical}, image edge detection  \cite{ding2001canny}, optics \cite{azadi2017statistical}, biomedical engineering \cite{kohler2003qrs}, neurophysiology \cite{petrantonakis2010emotion} seismology \cite{hancock2005effective}, machine learning \cite{huang2009frog}, and speech recognition and processing \cite{alias2016review,panagiotakis2005speech,park2009spatial}.

{\it Acknowledgment.} 
LL and TA acknowledges the Swedish Research Council  for funding (grant no. 2012-4526 and 2014-4305, respectively).

\bibliographystyle{unsrt} 



\section{Appendix}

\section{Probability that an odd number of zero-crossings occur}
In the main text we gave the probability $\omega(n)$ for an odd number of zero-crossings to occur, see Eq. \eqref{eq:omegan}. To determine $\omega(n)$ we start with the conditional probability that the number of zero-crossings $m$ is odd given that $X(n=0)=x_0>0$,
\begin{equation} \label{eq:om1}
\omega(n|x_0>0)=\int_{-\infty}^0 dx_nP (x_n|x_0>0),
\end{equation}
where $P (x_n|x_0>0)$ is the conditional probability density function (PDF) that the Brownian particle is at $x$ at time $n$ given the initial position $x_0>0$. It is given by \cite{brockwell2006introduction}
\begin{equation} \label{eq:GSP+cond+pdf}
P(x,n|x_0)=\frac{\text{exp}\left( -\frac{(x-x_0A(n))^2}{2\gamma(0)(1-A(n)^2)}  \right)}{\sqrt{2\pi\gamma(0)(1-A(n)^2)}},
\end{equation}
where $\gamma(n)=\langle x(n+k)x(k) \rangle$ is the covariance and $A(n)=\gamma(n)/\gamma(0)$ is the (normalized) autocorrelator. The stationary limit is
\begin{equation} \label{eq:stat+pdf}
\varrho(x)=\frac{1}{\sqrt{2\pi\gamma(0)}}\text{exp}\left( -\frac{x^2}{2\gamma(0)}  \right),
\end{equation}
since $A(n\to\infty)\to0$. The derivation of the conditional PDF is done in the following steps. Consider the covariance matrix 
 \[
   C=
  \left( {\begin{array}{cc}
   \gamma(0) & \gamma(n) \\
   \gamma(n) & \gamma(0) \\
  \end{array} } \right),
\]
where $C_{00}=C_{nn}=\gamma(0)$ and $C_{n0}=C_{0n}=\gamma(n)$. The joint PDF is given by,
\begin{equation}
P(x_n,x_0)=\frac{1}{2\pi\sqrt{\text{det}(C)}}\text{exp}\left( -\frac{1}{2}\sum_{j,k=\{0,n\}}x_jC^{-1}_{jk}x_k \right),
\end{equation}
where the inverse covariance matrix reads
 \[
   C^{-1}=
  \left( {\begin{array}{cc}
   \frac{1}{\gamma(0)(1-A(n)^2)} & -\frac{A(n)}{\gamma(0)(1-A(n)^2)} \\
   -\frac{A(n)}{\gamma(0)(1-A(n)^2)} & \frac{1}{\gamma(0)(1-A(n)^2)} \\
  \end{array} } \right).
\]
Thus,
\begin{equation}
P(x_n,x_0)=\frac{\text{exp}\left( -\frac{x_n^2-2x_nx_0A(n)+x_0^2}{2\gamma(0)(1-A(n)^2)}\right)}{2\pi\gamma(0)\sqrt{(1-A(n)^2)}}.
\end{equation}
Using Bayes' rule we get the conditional PDF via,
\begin{equation}
P(x_n|x_0)=\frac{P(x_n,x_0)}{\varrho(x_0)},
\end{equation}
which leads to Eq. \eqref{eq:GSP+cond+pdf}. 

Since we are considering stationary processes, the initial position is thermal and one must average Eq. \eqref{eq:om1} over the initial distribution that is stationary and given by Eq. \eqref{eq:stat+pdf}. If one has the initial position $x_0<0$ then an odd number of zero--crossings would correspond to $x(n)>0$. In our case it does not matter if $x_0$ is greater or less than zero. There will be as many zero--crossings in either case. Thus, we choose $x_0>0$ without loss of generality and find,
\begin{equation} \label{eq:om2}
\begin{aligned}
\omega(n)&=\int_{-\infty}^{\infty} dx_0\varrho (x_0) \omega(n|x_0>0) \\
&=\frac 1 2  + \frac{\arcsin A(n)}{\pi},
\end{aligned}
\end{equation}
where $x_0>0$ should be interpreted as $|x_0|$. 
\section{Formal inversion of the second moment}
In the main text, write $\langle m^2(z) \rangle$ in Eq. \eqref{eq:moments} using Eq. \eqref{eq:q_z} as
\begin{equation} \label{eq:rice6}
\langle m^2(z) \rangle \omega(z)= \frac{r^2z^2}{(z-1)^4}.
\end{equation}
Identify that products in $z$-space are due to convolutions and that $\mathcal{Z}^{-1}\left\{ \frac{z^2}{(z-1)^{4}} \right\}=\frac{1}{6}n(n^2-1)$ \cite{rade2013mathematics}, therefore
\begin{equation} \label{eq:rice7}
\sum_{k=0}^n\langle m^2(k) \rangle \omega(n-k)= \frac{r^2}{6}(n-1)n(n+1).
\end{equation}
Using that $\omega(0)=0$ and that $\omega(1)=r$ we pull out the term $\langle m^2(n) \rangle$ from the sum which yields the recursive formula
\begin{equation} \label{eq:rice8}
\langle m^2(n) \rangle=\frac{r}{6}n(n+1)(n+2)-\frac{1}{r}\sum_{k=0}^{n-1}\langle m^2(k) \rangle \omega(n-k+1).
\end{equation}
Higher order moments follows similarly, and below we give the recursive formula for the sub-leading part of the third moment.

\section{Expressions used in recursion relation for the sub-leading behavior}

The recursion equation for the sub-leading terms for the third moment is (compare to Eq.  (\ref{eq:second_moment_recur}) in the main text): 

\begin{eqnarray}\label{eq:third_moment_recur}
&&\langle m^3 (n) \rangle_s = {\cal G}^{(3)}(n)-\frac{1}{r^2}\sum_{k=0}^{n-1}\left(  \frac{rk}{2} +\langle m^3 (k) \rangle_s \right)\nonumber\\
&&\ \ \ \ \times\sum_{j=0}^{n+2-k}\omega(j)\omega(n-k-j+2),\nonumber\\
&&{\cal G}^{(3)}(n)=\frac{rn}{2}\left(  \frac{(n+1)(n+2)(n+3)(n+4)}{40}-2  \right)\nonumber\\
&&-r^3\Big[n(n-1) (n-2)-6 (n-1) n q^{(1)}(1)-24 q^{(1)}(1)^3\nonumber\\
&&+18 q^{(1)}(1) q^{(2)}(1)+3 n \left(6 q^{(1)}(1)^2-2 q^{(2)}(1)\right)-2 q^{(3)}(1)\Big]  \nonumber\\
&&-r\sum_{k=0}^{n-1}\Big\{  \frac{rk}{2}+  r^3\big[k(k-1) (k-2)-6 (k-1) k q^{(1)}(1)\nonumber\\
&&-24 q^{(1)}(1)^3+18 q^{(1)}(1) q^{(2)}(1)+3 k \big(6 q^{(1)}(1)^2-2 q^{(2)}(1)\big)\nonumber\\
&&-2 q^{(3)}(1)\big]   \Big\}\times\sum_{j=0}^{n+2-k}\omega(j)\omega(n-k-j+2) 
\end{eqnarray}

\section{Autocorrelator autoregressive process of order p}
The equation of motion of the AR($p$) process is given by \cite{brockwell2006introduction}
\begin{equation}
\begin{aligned}
x(n)=\sum_{k=1}^p\phi_kx(n-k)+\eta(n).
\end{aligned}
\end{equation}
The derivation of the corresponding autocorrelator starts by defining the backward operator $B$ as: $B^jx(n)=x(n-j)$. Let 
\begin{equation} \label{eq:match1}
\begin{aligned}
\phi(B)=1-\phi_1B-\phi_2B^2-\ldots -\phi_pB^p,
\end{aligned}
\end{equation}
then the equation of motion can be written on the compact form
\begin{equation}
\begin{aligned}
x(n)=\eta(n)/\phi(B).
\end{aligned}
\end{equation}
Next we set 
\begin{equation} \label{eq:match2}
\begin{aligned}
\phi(B)=(1-\lambda_1B)(1-\lambda_2B)\cdots (1-\lambda_pB),
\end{aligned}
\end{equation}
such that by equating like powers of $B$ in Eqs. \eqref{eq:match1} and \eqref{eq:match2}, each $\phi_k$ can be expressed in terms of the $\lambda$'s. Actually, for each $\lambda=1$, the equation of motion reduces to $\Delta^px(n)=\eta(n)$, where $\Delta$ is a discrete derivative satisfying $\Delta x(n)=x(n)-x(n-1)$. We make use of partial fraction that gives
\begin{equation} \label{eq:match3}
\begin{aligned}
\frac{1}{\phi(B)}=\frac{c_1}{1-\lambda_1B}+\frac{c_2}{1-\lambda_2B}\ldots +\frac{c_p}{1-\lambda_pB}.
\end{aligned}
\end{equation}
By writing on common denominator and match coefficients of like powers of $B$ (in the numerators of the LHS and RHS of Eq. \eqref{eq:match3}), one can show that 
the constants can be written
\begin{equation} \label{eq:const}
\begin{aligned}
c_k=\frac{\lambda_k^{p-1}}{\prod_{j\neq k}(\lambda_k-\lambda_j)}.
\end{aligned}
\end{equation}
Thus, for $|\lambda_k|<1$ the equation of motion becomes after expansion
\begin{equation}
\begin{aligned}
x(n)=\sum_{i=0}^{\infty}\sum_{k=1}^pc_k\lambda_k^i\eta(n-i),
\end{aligned}
\end{equation}
where we used that $B^i\eta(n)=\eta(n-i)$. Taking the expectation value $\gamma(n)=\langle\, x(n+k)x(k) \,\rangle$ and using that the noise $\eta(n)$ is Kronecker delta--correlated, $\langle \eta(n_1)\eta(n_2) \rangle=\Sigma^2\delta_{n_1,n_2}$, gives
\begin{equation} 
\begin{aligned}
\frac{\gamma(n)}{\Sigma^2}&=c_1\lambda_1^n\left[ \frac{c_1}{1-\lambda_1^2}+\frac{c_2}{1-\lambda_1\lambda_2}+\ldots \frac{c_p}{1-\lambda_1\lambda_p} \right] \\
&+c_2\lambda_2^n\left[ \frac{c_1}{1-\lambda_2\lambda_1}+\frac{c_2}{1-\lambda_2^2}+\ldots\frac{c_p}{1-\lambda_2\lambda_p} \right] \\
&+\ldots \\
&+c_p\lambda_p^n\left[ \frac{c_1}{1-\lambda_p\lambda_1}+\frac{c_2}{1-\lambda_p\lambda_2}+\ldots\frac{c_p}{1-\lambda_p^2} \right],
\end{aligned}
\end{equation}
which yields the autocorrelator via $A(n)=\gamma(n)/\gamma(0)$. 
%
\section{Asymptotic's of the moments}
We will need the following quantities to evaluate the asymptotic behavior of the first four moments:
\begin{align}
q(z) &\equiv \frac{2(z-1)\omega(z)}{z} \\
&=1-\frac{2(z-1)}{\pi}\sum_{n=0}^{\infty}\sin^{-1}(A(n))z^{-(n+1)} \\
q(1)&=1 \\
q^{(1)}(1)&=-\frac{2}{\pi}\sum_{n=0}^{\infty}\sin^{-1}(A(n)) \\
q^{(2)}(1)&=\frac{4}{\pi}\sum_{n=0}^{\infty}(n+1)\sin^{-1}(A(n)) \\
q^{(3)}(1)&=-\frac{6}{\pi}\sum_{n=0}^{\infty}(n+1)(n+2)\sin^{-1}(A(n)) \\
q^{(4)}(1)&=\frac{8}{\pi}\sum_{n=0}^{\infty}(n+1)(n+2)(n+3)\sin^{-1}(A(n))
\end{align}
%

Using the residue theorem for complex variables, see Eq. \eqref{eq:residue+asymptotic} in the main text, we calculate the asymptotic behavior of the first four moments from the pole at $z=1$.
\subsection{Second moment}
\begin{equation}
\langle m^2(z) \rangle = \frac{2r^2z}{(z-1)^3q(z)}
\end{equation}
Pole of order three at $z=1$:
\begin{equation}
\begin{aligned}
\langle m^2(n) \rangle &\sim \lim_{z\to1}\frac{1}{2!}\left( \frac{d}{dz} \right)^2 (z-1)^3\frac{2r^2z^n}{(z-1)^3q(z)} \\
&=\lim_{z\to1}r^2\left( \frac{d}{dz} \right)^2 \frac{z^n}{q(z)} \\
&=r^2\left(n(n-1)-2 n q^{(1)}(1)+2 q^{(1)}(1)^2-q^{(2)}(1)\right)
\end{aligned}
\end{equation}
%

\subsection{Third moment}
\begin{equation}
\langle m^3(z)\rangle = \frac{6r^3z}{(z-1)^4q(z)^2}-\underbrace{\frac{rz}{2(z-1)^2}}_{=Z\{rn/2\}}
\end{equation}
Pole of order four and two at $z=1$:
\begin{widetext}
\begin{equation}
\begin{aligned}
\langle m^3(n)\rangle &\sim \lim_{z\to1}\frac{1}{3!}\left( \frac{d}{dz} \right)^3 (z-1)^4\frac{6r^3z^n}{(z-1)^4q(z)^2}-\frac{rn}{2} \\
&=\lim_{z\to1}r^3\left( \frac{d}{dz} \right)^3 \frac{z^n}{q(z)^2}-\frac{rn}{2} \\
&=r^3\left[n(n-1) (n-2)-6 (n-1) n q^{(1)}(1)-24 q^{(1)}(1)^3\right.\\
&+\left.18 q^{(1)}(1) q^{(2)}(1)+3 n \left(6 q^{(1)}(1)^2-2 q^{(2)}(1)\right)-2 q^{(3)}(1)\right]-\frac{rn}{2}
\end{aligned}
\end{equation}
\end{widetext}

\subsection{Fourth moment}
\begin{equation}
\langle m^4(z)\rangle = \frac{24r^4z}{(z-1)^5q(z)^3}-\frac{4r^2z}{(z-1)^3q(z)}
\end{equation}
Pole of order five and three at $z=1$:
\begin{widetext}
\begin{equation}
\begin{aligned}
\langle m^4(n)\rangle &\sim \lim_{z\to1}\frac{1}{4!}\left( \frac{d}{dz} \right)^4 (z-1)^5\frac{24r^4z^n}{(z-1)^5q(z)^3}- \lim_{z\to1}\frac{1}{2!}\left( \frac{d}{dz} \right)^2(z-1)^3\frac{4r^2z^n}{(z-1)^3q(z)} \\
&= \lim_{z\to1}r^4\left( \frac{d}{dz} \right)^4 \frac{z^n}{q(z)^3}- 2\underbrace{\lim_{z\to1}r^2\left( \frac{d}{dz} \right)^2\frac{z^n}{q(z)}}_{\text{see }\langle m^2\rangle} \\
&=r^4\left\{n(n-1) (n-2) (n-3) -12 n(n-1) (n-2)  q^{(1)}(1)+360 q^{(1)}(1)^4\right.\\
&-360 q^{(1)}(1)^2 q^{(2)}(1)+36 q^{(2)}(1)^2+6 n(n-1) \left(12 q^{(1)}(1)^2-3 q^{(2)}(1)\right)+48 q^{(1)}(1) q^{(3)}(1)\\
&\left.+4 n \left(-60 q^{(1)}(1)^3+36 q^{(1)}(1) q^{(2)}(1)-3 q^{(3)}(1)\right)-3 q^{(4)}(1)\right\} -2\langle m^2(n)\rangle
\end{aligned}
\end{equation}
\end{widetext}

\end{document}